\documentclass[epj]{webofc}
\usepackage[varg]{txfonts}   % Web of Conferences font
\usepackage{epstopdf}
\woctitle{QUARKS-2016. 
          19th International Seminar on High Energy Physics}
\begin{document}
\title{Saturation effects in low-x DIS structure functions 
       and related hadronic total cross sections}

\author{\firstname{Francesco Giovanni} 
         \lastname{Celiberto}
         \inst{1,2}
         \fnsep
         \thanks{
          \email{francescogiovanni.celiberto@fis.unical.it}} 
        \and
        \firstname{Laszlo} 
         \lastname{Jenkovszky}
         \inst{3}
         \fnsep
         \thanks{\email{jenk@bitp.kiev.ua}} 
        \and
        \firstname{Volodymyr} 
        \lastname{Myronenko}
        \inst{4}
        \fnsep
        \thanks{\email{volodymyr.myronenko@desy.de}}
}

\institute{
Dipartimento di Fisica, Universit{\`a} della Calabria, 
Arcavacata di Rende, 87036 Cosenza, Italy
\and
Istituto Nazionale di Fisica Nucleare, 
Gruppo Collegato di Cosenza, 
\\ Arcavacata di Rende, 87036 Cosenza, Italy
\and
Bogolyubov Institute for Theoretical Physics (BITP), 
Ukrainian National Academy of Sciences
14-b, \\ Metrologicheskaya str., Kiev, 03680, Ukraine
\and
Deutsches Elektronen-Synchrotron DESY, Hamburg, Germany
          }

\abstract{%

High-energy nucleon total cross sections are related to low-x DIS structure functions by using
the additive quark model.}
\maketitle
%
%\section{Introduction}
%\label{intro}
%\section{Section title}
%\label{sec-1}
 In the additive quark model, the hadron-hadron total cross section can be written as a product of the cross sections of the constituents, $\sigma_{qq}$ \cite{J, JS}, {\it e.g.}
 \begin{equation}
 \sigma(s)^t_{pp}=\sigma_{qq}[n_V+n_S(s)]^2,
 \end{equation}
where $n_V$ is the number of valence quarks and $n_S(s)$ is that of sea quarks, their number increasing with energy.
 
It was suggested in Refs. \cite{J, JS} that the increasing number of sea quarks is related to the
Bjorken scaling-violating contribution to the deep inelastic lepton-hadron structure function (DIS SF), namely to the momentum fraction of the relevant quarks given by the integral over the DIS structure function $F_2(x,Q^2)$. In Ref. \cite{J} a simple model for the DIS structure function, known at those times, was used, resulting in the following expression for the total cross section, compatible with the data
\begin{equation}
\sigma(s)^t_{pp}=\sigma_{qq}n_V^2(1+0.0.16 \, \ell n (s/Q^2_0)),
\end{equation}  
where $\sigma_{qq}$ is a free parameter, $Q^2_0$ was fitted to the DIS data, and $n_V=3$ . 

In Ref. \cite{JS} the DIS SF was related to hadronic cross sections by means of finite-energy sum rules in $Q^2$.   

The number of quarks in a reaction can be calculated from the SF by means of sum rules. see e.g. 
\cite{Close, Roberts}.

In Ref. \cite{D} following ansatz for the small-$x$ singlet part (labelled by the
upper index $S,0$) of the proton structure function, interpolating
between the soft (VMD, Pomeron) and hard (GLAP evolution) regimes was proposed:           

$$F_{2}^{(S,0)}(x,Q^2) =
A\left({Q^2\over Q^2+a}\right)^{1+\widetilde{\Delta} (Q^2)}
e^{\Delta (x,Q^2)}  ,                                             \eqno(3.1)$$
with the "effective power"
$$ \widetilde{\Delta }(Q^2) =\epsilon+\gamma_1\ell n {
\left(1+\gamma_2\ell n{\left[1+{Q^2\over Q^2_0}\right]}\right)} ,
                                                                  \eqno(3.2)$$
and
$$\Delta (x,Q^2) = \left(\widetilde{\Delta } (Q^2)  \ell n{x_0\over x}\right)
                   ^{f(Q^2)},                                     \eqno(3.3)$$
where
$$ f(Q^2) = {1\over 2}\left( {1+e^{-{Q^2/Q_1^2}}}\right) .        \eqno(3.4)$$

At small and moderate values of $Q^2$, the exponent $\widetilde{\Delta}(Q^2)$ (3.2) may be interpreted as a $Q^2$-dependent "effective Pomeron intercept", as shown in Fig. 1.

The function $f(Q^2)$ has been introduced in order to provide for the
transition from the Regge behaviour, where $f(Q^2)=1$, to the 
asymptotic solution of the GLAP evolution equation, where $f(Q^2)=1/2$. 

In Ref. \cite{D} the above singlet SF was appended by a non-singlet part, important at
large values of $x$. The parameters were fitted to the DIS data in a wide range of $x$ and $Q^2$.The values of the fitted parameters are: $A=0.1623,\ \ a=0.2916$ GeV$^2,\ \ 
\gamma_2=0.01936,\ \ Q_0^2=0.1887$ GeV$^2,\ \ Q_1^2=916.1$ GeV$^2;\ \  x_0=1,\ \ \epsilon=0.08,\ \ \gamma_1+2.4$ were fixed (by QCD-related arguments). The resulting fits and more details can be found in Ref. \cite{D}. 

\begin{figure}
  \centering
  \includegraphics*[width=7cm,height=7cm]{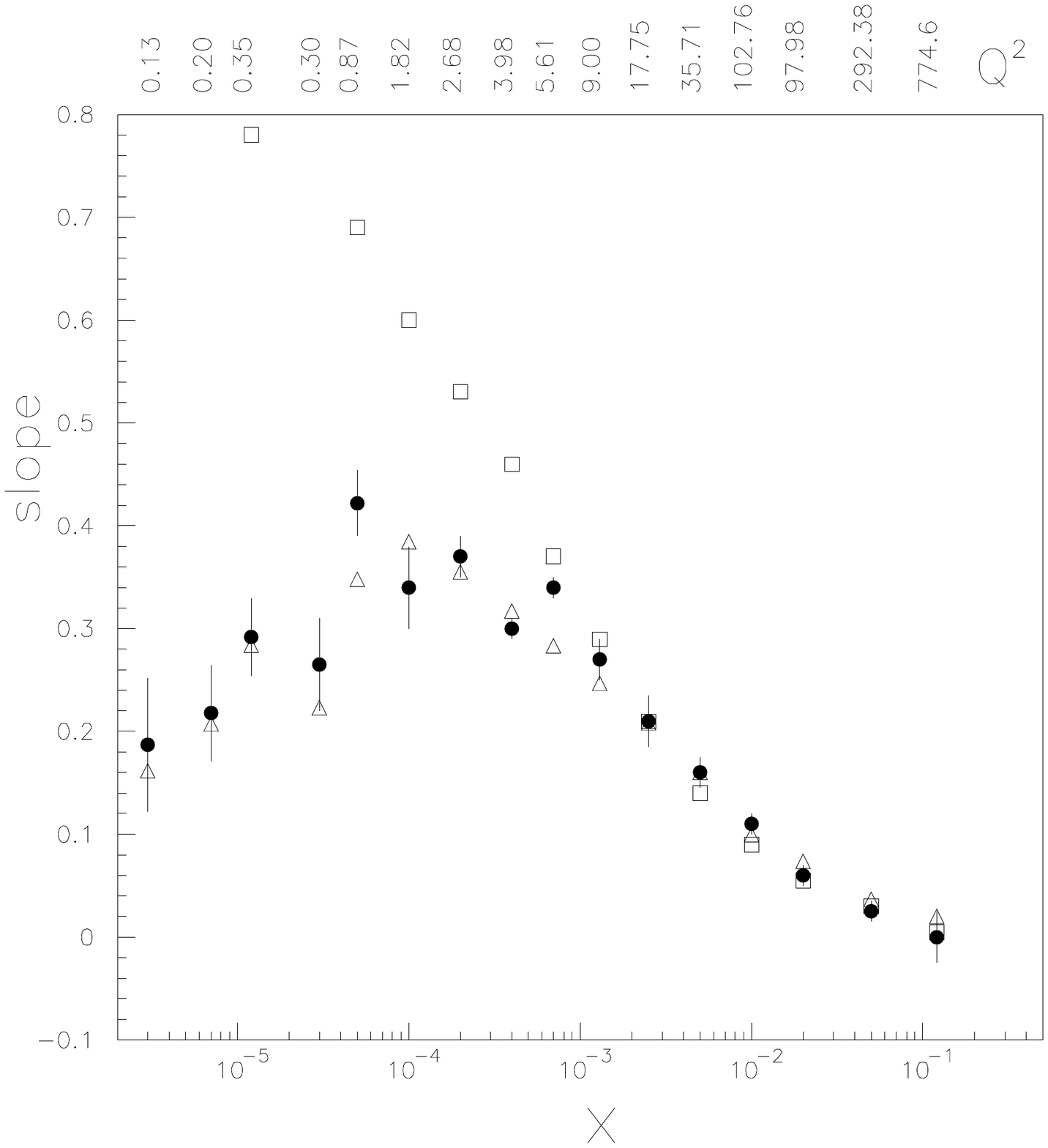}\hspace*{.1cm}
  \includegraphics*[width=7cm,height=7cm]{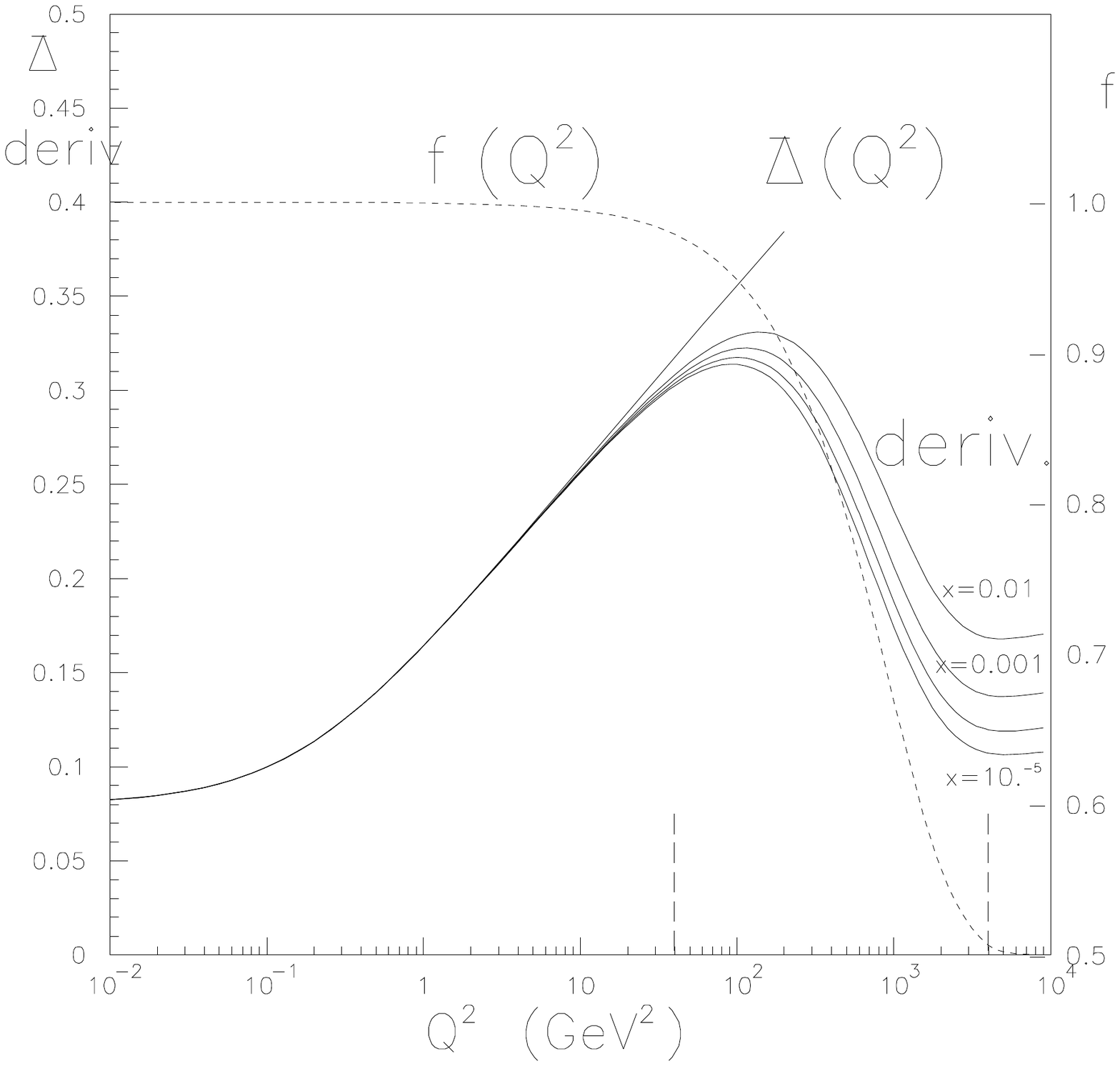}
  \caption{Slope of the structure function $F_2(x,Q^2)$.}
  \label{fig:res_DPM_ePfixed}
 \end{figure}
  
The proton-proton total cross section is cast by integrating Eqs. (3) between $x=0$ and $x=1$. At high energies, only the singlet part of the SF, Eqs. (3) (the "Pomeron") is relevant. Integration can be performed numerically. The result is in reasonable agreement with the data on $pp$ total cross sections, including those from the LHC.  

\begin{figure}[tbp]
\vspace{-0.3cm} 
%%\vspace*{5pt}
\centerline{
\includegraphics[scale=0.35]{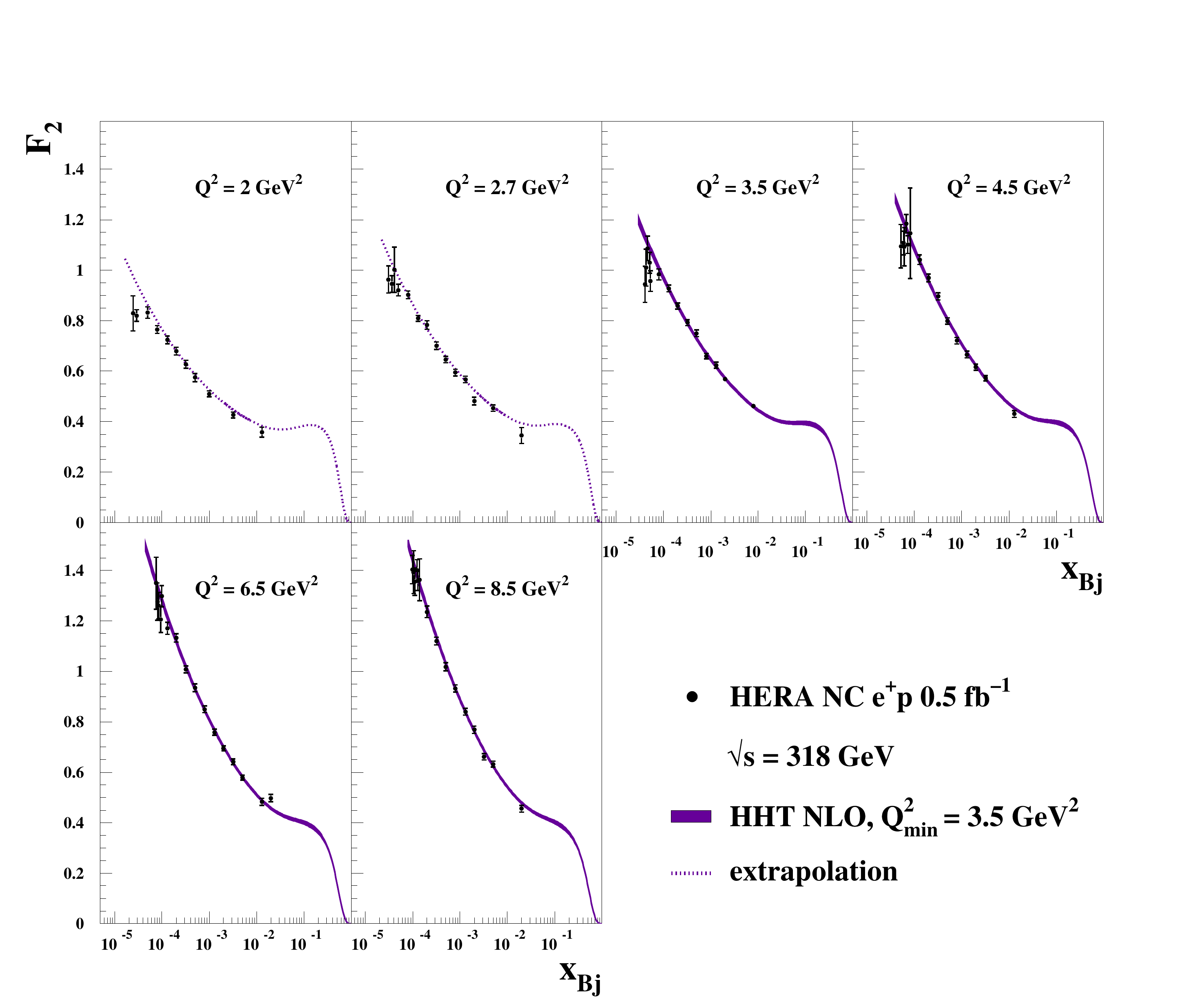}}
\centerline{
\includegraphics[scale=0.35]{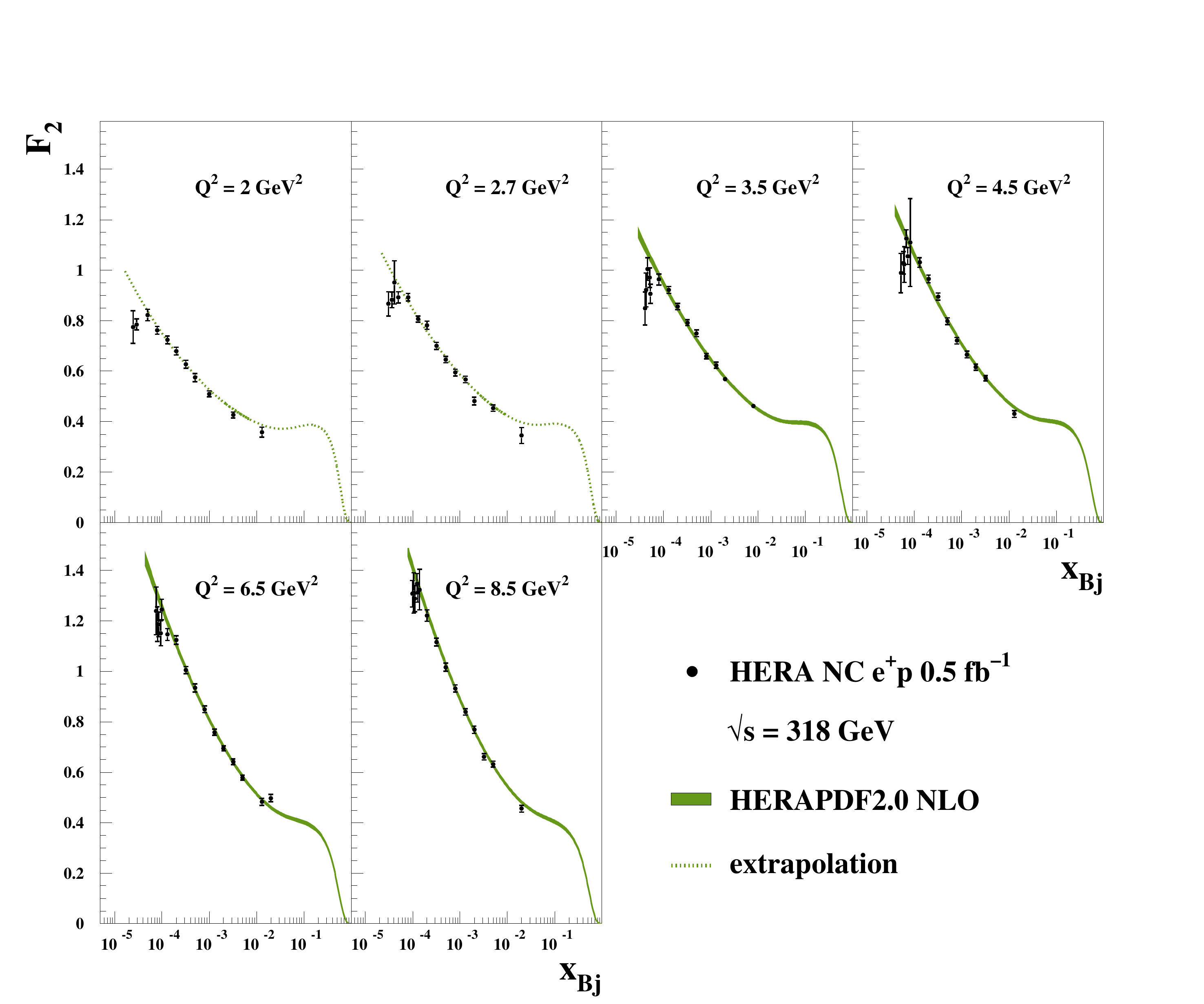}}
\vspace{0.5cm}
\caption {HHT (top) and HERAPDF2.0 (bottom) NLO predictions 
          of the $F_2$ structure function 
          at $Q^2_{\rm min} = $~$3.5$\,GeV$^2$,
          compared to extracted values. 
          For more details, see Ref.~\cite{Abt:2016vjh}.
}
\label{fig:f2htq2gt2}
\end{figure}

The operator-product expansion beyond leading twist has diagrams in which two, three or four
gluons may be exchanged in the $t$-channel such that these gluons may be viewed as recombining.
This recombination could lead to gluon saturation. 
The colour-dipole framework also inspired a phenomenological model of saturation by
Golec--Biernat--W\"usthoff (GBW), in which the onset of saturation is characterised as the transition
from a ``soft'' to a ``hard'' scattering regime. This occurs along a ``critical line'' in the $x_{Bj},\ \  Q^2$ plane. 

Recently new results on low-$x$ DIS parton distrubutions (PDF and SF) have appeared \cite{Abramowicz:2015mha, Abt:2016vjh}; they show an intriguing change of regime towards smallest values of $x$  (Fig.~\ref{fig:f2htq2gt2}) - possible saturation effect? 

We intend to investigate its impact on the asymptotic behaviour of the total cross sections.   
  
\subsection*{Acknowledgement}
\label{Acknowledgement}
L.J. thanks the Organizers of the Quarks2016 conference for their warm hospitality and support. 

%
% BibTeX or Biber users please use 
% (the style is already called in the class, 
%ensure that the "woc.bst" style is in your local directory)
% \bibliography{name or your bibliography database}

\end{document}